\documentclass[12pt]{article}

\usepackage{physics}
\usepackage{enumerate}
\usepackage{empheq}
\usepackage{booktabs}
\usepackage{amsmath,amssymb,amsbsy,amstext, amsthm, simplewick}
\usepackage{hyperref}
\usepackage{graphicx}
\usepackage{amsfonts}
\usepackage{mathrsfs}
\usepackage{color}
\usepackage{framed}
\usepackage{wasysym}
\usepackage{cite}
\usepackage[utf8]{inputenc}
\usepackage{tikz-feynman}

\newcommand{\Comment}[1]{{}}
\definecolor{MyDarkBlue}{rgb}{0.15,0.15,0.45}

\newcommand{\D}{{\rm d}}

\newcommand{\Mpl}{M_{\text{Pl}}}
\newcommand{\ct}{c_{\text{T}}}

\title{Nonrenormalization in beyond-Horndeski}
\author{}
{

\begin{document}

\begin{center}
{\Large\bf{Behind Horndeski:\\[0.3cm]Structurally Robust Higher Derivative EFTs}}
\end{center} 

 \vspace{0.5truecm}
\thispagestyle{empty} 
\centerline{
Luca Santoni${}^{a,}$\footnote{E-mail address: l.santoni@uu.nl},
Enrico Trincherini${}^{b,c,}$\footnote{E-mail address: enrico.trincherini@sns.it},
Leonardo G. Trombetta${}^{b,c,}$\footnote{E-mail address: leonardo.trombetta@sns.it}
}

\vspace{0.5 cm}

\centerline{\it $^a$Institute for Theoretical Physics and Center for Extreme Matter and Emergent Phenomena,}
\centerline{\it Utrecht University, Leuvenlaan 4, 3584 CE Utrecht, The Netherlands}

 \vspace{.1cm}

\centerline{\it $^b$Scuola Normale Superiore, Piazza dei Cavalieri 7, 56126, Pisa, Italy}

\vspace{.1cm}

\centerline{\it $^c$INFN - Sezione di Pisa, 56100 Pisa, Italy}

\vspace{2cm}
\begin{abstract}

Higher derivative scalar interactions can give rise to interesting cosmological scenarios. We present a complete classification of such operators that can yield sizeable effects without introducing ghosts and, at the same time, define an effective field theory robust under the inclusion of quantum corrections. A set of rules to power count consistently the coefficients of the resulting Lagrangian is provided by the presence of an approximate global symmetry. The interactions that we derive in this way contain a subset of the so-called Horndeski and beyond Horndeski theories. Our construction therefore provides a structurally robust context to study their phenomenology. Applications to dark energy/modified gravity and geodesically complete cosmologies are briefly discussed.   

\end{abstract}

\newpage

\thispagestyle{empty}
\tableofcontents
\newpage
\setcounter{page}{1}
\setcounter{footnote}{0}

\section{Introduction}

The theory of cosmic inflation is commonly considered the most successful
framework to describe the evolution of our Universe at early times. Consisting in an accelerated expansion, a microscopic description of its origin is nevertheless still lacking so far. Few billions years after that epoch, the Universe is now undergoing an equally mysterious phase of acceleration, driven by what is generically known as Dark Energy (DE). A great deal of attention has been devoted, in both cases, to the analysis of scalar theories coupled to gravity as prototypical examples of the dynamics behind acceleration. The result is a glut of different models, proposed over the years in the literature. 

In the absence of a `best motivated' proposal for the dynamics of the new degree of freedom, however, a convenient choice to characterize the phenomenology of the different models is to follow an Effective Field Theory (EFT) approach. Assuming that the EFT for the scalar field $\phi$ is valid up to a UV cutoff scale $\Lambda$, one needs a hypothesis on the typical size of the infinite tower of higher dimensional operators that appear in the EFT Lagrangian. Such an estimate for the operators is called the `power counting rule' of the effective theory. For example, the simplest assumption is that the theory of $\phi$, apart from $\Lambda$, is characterized only by a coupling $g$. In this case the Lagrangian will be of the form:
\begin{equation}
\label{eftpc}
 \mathcal{L} = \frac{\Lambda^4}{g^2} L \left( \frac{\partial}{\Lambda} , \frac{g \phi}{\Lambda} \right),
\end{equation} 
where $L$ is a function with $\mathcal{O} (1)$ dimensionless parameters. This power counting has been, more or less explicitly, at the basis of most of the models of Inflation and Dark Energy since their emergence. In this case, the effect of higher derivative (HD) operators, i.e. with more than one derivative per field, is such that they either give subleading contributions to physical observables or the derivative expansion breaks down and the EFT in (\ref{eftpc}) is no longer a good description. In particular, this means that the scale of the ghost-like instability usually associated with HD is at or above the cutoff and therefore the theory is perfectly well-defined at energies below $\Lambda$. This is what happens, for example, in the case of the QCD chiral Lagrangian, which describes the dynamics of pions below $\Lambda_{\rm QCD}$ and contains, as every effective field theory, infinitely many HD operators.     

In search for models for the accelerated expansion of the Universe that differ qualitatively from the ones described by the EFT in (\ref{eftpc}),  
in the last decade there has been a considerable interest in scenarios where the simple energy expansion in (\ref{eftpc}) is modified in such a way that HD operators become at least as important as the one with less derivatives, within the domain of validity of the low energy theory. This feature should be made robust by the presence of some symmetry, exact or approximate, that can provide, at least in principle, a different set of rules to power count the coefficients of the effective Lagrangian, even in the absence of an explicit UV completion.      

A necessary condition to fulfill this goal is that the HD operators providing the leading contribution to physical observables should not introduce any instability within the regime of validity of the EFT. The simplest example of this situation is perhaps given by the {\it galileon} EFT \cite{Nicolis:2008in}: the invariance under the galileon (plus shift) symmetry 
\begin{equation}
\label{galileonsym}
\phi \to \phi + c + b_\mu x^\mu
\end{equation} 
guarantees that the equations of motion obtained from the three leading HD operators are of second order. 
This example motivated a lot of activity to find the most general HD interactions of a scalar coupled to gravity that do not introduce any additional, and necessarily ghost-like, degree of freedom. 

On the other hand, as we previously introduced, a second necessary ingredient is a power counting rule. It allows to estimate, also when quantum corrections are included, if there is a finite number of operators---or a symmetry that relates their coefficients\footnote{An example in the context of shift symmetric theories that include at the leading order only single derivatives of the scalar field---generically known as $P(X)$ theories, where $X$ is defined later on in \eqref{defXY}---is provided by the DBI action \cite{Silverstein:2003hf,Alishahiha:2004eh}. Here, the explicit form of the function $P(X)$ is dictated by a non-linearly realized higher-dimensional spacetime symmetry, which results in very specific relations among the coefficients of the different Lagrangian operators.}---that affects the observable of interest or if, instead, the result is uncalculable, because there are infinitely many contributions of the same order. In the case of the galileon it is precisely the symmetry (\ref{galileonsym}) that controls the structure of the operators: the ones with at most one derivative per field are not generated by quantum corrections and the leading HD terms, which are finite in number (three), are also not renormalized \cite{Luty:2003vm}.  

However, invariance under galileon transformations cannot be exact in any application to cosmology because every coupling of the field to gravity breaks it explicitly. A more structured theory is therefore needed: following what we proposed in \cite{Pirtskhalava:2015nla}, the transformations in Eq.~\eqref{galileonsym} can guide the formulation of a power counting for the scalar field characterized by two energy scales: the cutoff, that we will denote in the paper as $\Lambda_3$, and a second scale $\Lambda_2 \equiv ( M_{\rm Pl} \Lambda^3_3)^{1/4}$, parametrically higher assuming $\Lambda_3 \ll M_{\rm Pl}$, that controls the breaking of the galileon symmetry\footnote{From now on we are assuming that couplings are $O(1)$ and we do not write them explicitly; we also omit factors of $4\pi$ for simplicity.}.    

To find all the theories that conform to this new power counting, we will search for the most general set of interactions of the scalar to gravity that generate galileon symmetry breaking quantum corrections suppressed by the highest possible scale. As we will see in detail, this guarantees that:  \textit{i)} in the resulting theory higher derivative interactions are at least as important as the others on interesting cosmological backgrounds and \textit{ii)} the leading operators, in principle infinite in number, receive only small quantum corrections, suppressed by integer powers of $(\Lambda_3/ \Lambda_2)^{4}$, and therefore it is technically natural to introduce only a finite number of them, in such a way that physical observables can be reliably computed.   

The condition that we impose on the interactions is powerful enough that allows us to re-derive, though from a very different perspective, both the Horndeski \cite{Horndeski:1974wa, Deffayet:2011gz} and the so called beyond Horndeski \cite{Gleyzes:2014dya,Zumalacarregui:2013pma} theories. Or, to be more precise, the most general subset of the two that enjoys the properties \textit{i)} and \textit{ii)} defined in the previous paragraph. Our results generalize what we obtained previously in \cite{Pirtskhalava:2015nla} and hence we continue to refer to this more general class as `theories with weakly broken galileon (WBG) invariance'.  

There are several phenomenological applications that make HD EFTs particularly interesting.
We will briefly touch two of them: dark energy/modify gravity and early Universe cosmologies alternative to inflation. In the former case, the robustness of WBG theories is used to infer the naturalness of an exactly luminal speed of propagation for gravitational waves around the medium that gives rise to the accelerated expansion \cite{Creminelli:2017sry,Baker:2017hug,Ezquiaga:2017ekz,Sakstein:2017xjx}, while in the latter we emphasize how a particular class of HD theories can be used to construct geodesically complete cosmologies that are stable along the whole evolutionary trajectory \cite{Creminelli:2016zwa,Cai:2016thi}.

The paper is organized as follows. In Sec.~\ref{sec:wbgpc} we discuss the power counting rule for theories where HD interactions satisfy the condition \textit{i)}. Then, in Sec.~\ref{completeWBG} we find the most general theory up to quadratic order in second derivatives enjoying WBG invariance and thereby satisfying also condition \textit{ii)}. We achieve this by demanding that all possibly dangerous quantum corrections cancel. This is the main result of our work. In Sec.~\ref{sec:HandBH} we relate the class of interactions obtained in this way to the well known Horndeski and beyond-Horndeski theories. Afterwards, Sec.~\ref{sec:pheno} is devoted to the discussion of the applications to the EFT of DE and to geodesically complete cosmologies. Finally, in Sec.~\ref{sec:concl} we present our conclusions and outlook. The Appendices provide further details on the calculations.

\section{The WBG power counting}
\label{sec:wbgpc}

The main goal of the paper is to identify a class of models where higher derivative interactions play a crucial role. As we discussed in the introduction, the simplest example of a robust theory satisfying this requirement is given by a Lagrangian that enjoys the symmetry (\ref{galileonsym}). In particular, we are interested in the only two operators\footnote{There is one last operator that belongs to the same class, schematically of the form $(\partial \phi)^2 (\partial^2 \phi)^3$. However in the paper we will never include operators that are cubic in $(\partial^2 \phi)$. This is a consistent choice, as we discuss later in the section.}, together with the kinetic term, that are invariant up to a total derivative \cite{Nicolis:2008in}:
 \begin{equation}
 \label{galileon34}
\frac{  (\partial \phi)^2 \Box \phi}{\Lambda_3^3} , \; \frac{  (\partial \phi)^2 ((\Box \phi)^2 - \partial_\mu  \partial_\nu \phi  \partial^\mu  \partial^\nu \phi) }{\Lambda_3^6}.
 \end{equation}
These are the leading ones in a derivative expansion---all the others are of the form $\partial^m (\partial^2 \phi)^n$--- and since their equation of motion (EOM) is second order they do not introduce the instability usually associated with HD operators, the so-called Ostrogradsky ghost. 
Therefore, their effect can become the dominant one, at least for some energy (or length) scales, within the regime of validity of the EFT.

Once the scalar field $\phi$ is coupled to gravity, minimally or not, the galileon symmetry is inevitably broken (we will instead always assume in the paper that the shift symmetry, $\pi \to \pi + c$, is preserved\footnote{For the physical implications of an internal shift symmetry in the context of FLRW cosmologies we refer to \cite{Finelli:2017fml,Finelli:2018upr}.}), in particular operators of the form
 \begin{equation}
 \label{PX}
\frac{(\partial \phi)^{2n}}{ \Lambda^{4n-4}_{n}} \; , 
 \end{equation}
with less derivative than the two galileon interactions, will be generated. If they were suppressed by the same scale $\Lambda_3$ as the ones in (\ref{galileon34}), they would give the dominant contribution and the theory would simply be a generic shift invariant model. We are instead looking for a theory where physical observables receive corrections from HD operators that are at least $O(1)$ compared to the standard scenario described by the power counting in Eq.~\eqref{eftpc}. It is then natural to ask what the smallest value for $\Lambda_{n}$ should be to obtain this result. The answer in general depends on the background solution one is interested in. Since the main application we have in mind for the HD EFT is to describe the late time or the early Universe accelerated expansion, we will consider an FLRW-type background metric and a time-dependent background $\phi_0(t)$ for the scalar. 

Let us now start with an estimate of $\Lambda_{n}$ for the first interaction, i.e. $n=2$, and for simplicity we include only the contribution of the second operator in Eq.~\eqref{galileon34}: the first Friedmann equation reads, schematically,
\begin{equation}
 H^2 = \frac{\rho}{3 \Mpl^2} \sim \frac{1}{\Mpl^2} \left( \dot{\phi}_0^2 + \frac{\dot{\phi}_0^4}{\Lambda_2^4} + \dot{\phi}_0^2 \frac{H^2 \dot{\phi}_0^2}{\Lambda_3^6} \right) \; ,
\end{equation}
where we have assumed that the background field satisfies $\ddot{\phi}_0 \ll H \dot{\phi}_0$, and thus $\square \phi_0 \sim H \dot{\phi}_0$, with $H$ the Hubble parameter.

We can see that, for the second and third terms to contribute to the energy density as much as the kinetic term, we must have both
\begin{equation}
X_0 \lesssim 1, \quad Y_0 \lesssim 1,
\end{equation}
on the background, where we are introducing the notation
\begin{equation}
X \equiv -\frac{\nabla_\mu\phi\nabla^\mu\phi}{\Lambda_2^4}, \quad Y \equiv \frac{\nabla_\mu\nabla^\mu \phi}{\Lambda_3^3}.
\label{defXY}
\end{equation}
This implies that the two scales satisfy $\Lambda_3^3 \sim H \Lambda_2^2$. Furthermore, solving the first Friedmann equation under this regime gives $H \sim \Lambda_2^2/\Mpl$. Using both relations to eliminate $H$, we conclude that the scale of the symmetry breaking operator should be 
\begin{equation}
\Lambda_2 \sim (\Mpl \Lambda_3^3)^{1/4}.
\label{lambda2}
\end{equation}

We can repeat the same analysis including operators with extra powers of $X$ and, in general, for arbitrary functions of $X$ in front of the $(\partial^2 \phi)^m$ factor. The conclusion is that all the operators satisfying the power counting
\begin{equation}
 \frac{(\nabla \phi)^{2n}}{\Lambda_2^{4(n-1)}} \frac{(\nabla \nabla \phi)^m}{\Lambda_3^{3m}},
 \label{Op}
\end{equation}
are equally important on the background we are considering.
The next question we need to address is whether the structure (\ref{Op}) is robust or, in other words, if the contributions that come from quantum corrections to those operators are at most of the order of the estimates that appear in (\ref{Op}).  
As we showed already in \cite{Pirtskhalava:2015nla}, the answer is in general negative. It is easy to check, for example, that the quartic galileon in Eq.~\eqref{galileon34}, minimally coupled to gravity, contains a vertex of the form: 
\begin{equation}
\frac{  (\partial \phi)^3  \partial^2 \phi \,  \partial h }{M_{\rm Pl}\Lambda_3^6}.
 \end{equation}
This interaction generates corrections of the order $(\nabla \phi)^{6a}/(M_{\rm Pl}^{2a} \Lambda_3^{10a-4})$ that are much larger, for $a>1,$ than the corresponding $n=3a$, $m=0$ operators in Eq.~\eqref{Op}. 
This example shows that a generic coupling to gravity completely spoils the non-renormalization properties associated with (unbroken) galileon invariance. When the symmetry is exact, indeed, operators with less than two derivatives per field are not generated at the quantum level \cite{Luty:2003vm}. In the presence of a generic explicit breaking not only those operators are obviously generated but $(\nabla \phi)^{6a}$ turn out to be the dominant ones and the resulting theory is, up to small corrections, just a particular case of a $P(X)$ Lagrangian.

It is clear at this point that a robust higher derivative EFT can contain only a very specific set of operators that scale as in Eq.~\eqref{Op}. If we label the elements of this group as ${\cal O}^\text{I}$ and their ${\cal O}(1)$ coefficients as $c^\text{I}$, they are defined by the property that the contributions generated by quantum corrections, $\delta c^\text{I}$, satisfy $\delta c^\text{I} \ll c^\text{I}$.

In \cite{Pirtskhalava:2015nla} we identified an example of such a subset of operators. We proved that they receive quantum corrections that are always suppressed by powers of $\frac{\Lambda_3}{\Mpl}$. 
Focussing for instance on the quantum mechanically generated operators $(\nabla \phi)^{2n}$, we have shown that at loop level they always scale as
\begin{equation}
\sim \frac{(\nabla \phi)^{2n}}{\Mpl^n \Lambda_3^{3n-4}} \, .
\label{qcH}
\end{equation}
The same result holds also for the other operators $(\nabla \phi)^{2n}(\nabla\nabla \phi)^m$ with $m\neq0$.
Thus, inheriting a remnant of the non-renormalization properties of the galileon, in \cite{Pirtskhalava:2015nla} we have dubbed the theories with this property `WBG theories'. Their phenomenology, in the context of inflation, has been studied afterwards in \cite{Pirtskhalava:2015zwa,Pirtskhalava:2015ebk}. In the next section, we generalize the proof of \cite{Pirtskhalava:2015nla} and find the most general class of WBG operators ${\cal O}^{\rm I}$, up to quadratic order in the second derivatives of the scalar field\footnote{One could in principle generalize our result to include in ${\cal O}^{\rm I}$ also operators that are cubic in the second derivatives of the scalar field (i.e. with $m=3$), but for the sake of simplicity of the presentation we decide not to do it here. Notice that setting them `to zero' in the Lagrangian does not yield fine tuning problems: indeed, as shown later on in Tab. \ref{pctable}, they are generated at a scale that is parametrically larger than the one suppressing the quadratic operators in ${\cal O}^{\rm I}$ and, for this reason,  they can be safely disregarded.}. 

Before getting to that, it is worth stressing that in a WBG theory, as in any genuine EFT, all kinds of interactions allowed by the symmetries are included. In particular, it means that there will be also operators, let us call them ${\cal O}^{\rm II}$, with the same number of fields and derivatives as ${\cal O}^{\rm I}$ but a different contraction of indices such that they do not enjoy any non-renormalization property. The point is that it is consistent to assume that they are suppressed by an additional factor $\frac{\Lambda_3}{\Mpl}$ at tree level compared to the WBG ones because this is precisely the size of the contributions they receive from quantum corrections. In other words, they follow a different power counting that reads:
\begin{equation}
\frac{(\nabla \phi)^{2n}}{\Lambda_2^{4n}} \frac{(\nabla \nabla \phi)^m}{\Lambda_3^{3m-4}} \, .
\label{Op2}
\end{equation}
For completeness, we recall that there is a third group ${\cal O}^{\rm III}$ of operators, already present in the effective theories, that contain at least two derivatives per field and are trivially generated at the scale $\Lambda_3$ (they are precisely the operators that become trivially galileon invariant at Lagrangian level on flat spacetimes).

In conclusion, the power counting of higher derivative EFT that are robust under the inclusion of quantum corrections is summarized in Table \ref{pctable}.

\begin{table}
\begin{center}
\begin{tabular}{| c | c | c |}\hline   
${\cal O}^\text{I} $& $ \frac{(\nabla \phi)^{2n}}{\Lambda_2^{4(n-1)}} \frac{(\nabla \nabla \phi)^m}{\Lambda_3^{3m}}$ & $\frac{ \delta c^\text{I}}{c^\text{I}} \sim \frac{\Lambda_3}{\Mpl}$ \\ [1ex]  \hline
${\cal O}^\text{II} $& $ \frac{(\nabla \phi)^{2n}}{\Lambda_2^{4n}} \frac{(\nabla \nabla \phi)^m}{\Lambda_3^{3m-4}} $ & $\frac{\delta c^\text{II}}{c^\text{II}} \sim {\cal O}(1)$ \\ [1ex]  \hline
${\cal O}^\text{III} $ & $\frac{\nabla^m(\nabla \nabla \phi)^n}{\Lambda_3^{3n+m-4}}$ & $ \frac{ \delta c^\text{III}}{c^\text{III}} \sim {\cal O}(1)$ \\ [1ex] \hline
\end{tabular} 
\end{center}
\caption{The power counting of higher derivative EFTs. The most general set of operators in group I is derived in Sec.~\ref{completeWBG}.}
\label{pctable}
\end{table}

\section{The most general WBG theory up to quadratic order in $\nabla\nabla\phi$}
\label{completeWBG}

In this section we explicitly construct the most general class of operators of type ${\cal O}^{\rm I}$ (see Tab. \ref{pctable}) up to quadratic order in the second derivatives of the scalar field.

Starting from the case of operators with $m=0$, it is straightforward to show that the shift symmetric Lagrangian
\begin{equation}
S_2 = \Lambda_2^4 \int \D^4 x\sqrt{-g} \,  G_2(X) \, ,
\label{S2}
\end{equation}
where $G_2$ is a function of $X$, defined in \eqref{defXY}, satisfies the scaling \eqref{qcH} at loop level and therefore the non-renormalization properties displayed in the corresponding row in Tab. \ref{pctable}. Then, it is by default entitled to be in the WBG class.

At the next-to-leading order we include operators linear in $\nabla\nabla\phi$, i.e. with $m=1$. The most general Lagrangian at this order can be written as \cite{Deffayet:2010qz}
\begin{equation}
S_3 = \Lambda_2^4 \int \D^4 x\sqrt{-g} \,  G_3(X) \frac{\square\phi}{\Lambda_3^3} \, .
\label{S3}
\end{equation}
Indeed, notice that any other combination involving the contraction $\nabla^\mu\phi\nabla^\nu\phi\nabla_{\mu}\nabla_{\nu}\phi$ can be easily recast in the general form \eqref{S3} after straightforward integrations by parts.
In \cite{Pirtskhalava:2015nla} it has been shown that loops involving vertices of type \eqref{S3} generate interactions at quantum level that scale as those in \eqref{qcH}, corresponding again to the small corrections $\delta c^\text{I}/c^\text{I}\sim \Lambda_3/\Mpl$, as displayed in Tab. \ref{pctable}.
We refer to \cite{Pirtskhalava:2015nla} for further comments on this point. 

Moving on to the case with $m=2$, the number of inequivalent contractions increases. Now the most general shift symmetric scalar-tensor theory (that depends quadratically on the second derivatives of the scalar field $\phi$) reads \cite{Achour:2016rkg}
\begin{equation}
S_4 = \int \D^4 x\sqrt{-g} \left[
\frac{\Lambda_2^8}{\Lambda_3^6} f(X)R  + \frac{\Lambda_2^4}{\Lambda_3^6} C^{\mu\nu,\rho\sigma} \nabla_\mu\nabla_\nu\phi\nabla_\rho\nabla_\sigma\phi 
\right] \, ,
\label{STT}
\end{equation}
where $C^{\mu\nu,\rho\sigma}$ is a tensor made of products of $\nabla_\mu\phi$ only. By construction, it can be always written in such a way to have the following symmetry structure
\begin{equation}
C^{\mu\nu,\rho\sigma} = C^{\nu\mu,\rho\sigma} = C^{\mu\nu,\sigma\rho} = C^{\rho\sigma, \mu\nu} \, .
\end{equation}
Then, the most general form of $C^{\mu\nu,\rho\sigma}$ is \cite{Achour:2016rkg}
\begin{multline}
C^{\mu\nu,\rho\sigma} = \frac{\alpha_1(X)}{2}\frac{(\nabla\phi)^2}{\Lambda_2^4} (g^{\mu\rho}g^{\nu\sigma}+g^{\mu\sigma}g^{\nu\rho}) + \alpha_2(X) \frac{(\nabla\phi)^2}{\Lambda_2^4} g^{\mu\nu}g^{\rho\sigma}
\\
+ \frac{\alpha_3(X)}{2\Lambda_2^4}\left(g^{\rho\sigma} \nabla^\mu\phi\nabla^\nu\phi  + g^{\mu\nu} \nabla^\rho\phi\nabla^\sigma\phi\right) 
\\
+ \frac{\alpha_4(X)}{4\Lambda_2^4} \left(g^{\nu\sigma} \nabla^\mu\phi\nabla^\rho\phi  + g^{\mu\sigma} \nabla^\nu\phi\nabla^\rho\phi + g^{\nu\rho} \nabla^\mu\phi\nabla^\sigma\phi  + g^{\mu\rho} \nabla^\nu\phi\nabla^\sigma\phi\right)   
\\
+ \frac{\alpha_5(X)}{\Lambda_2^8}\nabla^\mu \phi\nabla^\nu \phi\nabla^\rho \phi\nabla^\sigma \phi \, ,
\label{STT-2}
\end{multline}
where $\alpha_i$ are arbitrary functions of $X$.

The scales in \eqref{STT} have been chosen in such a way that the operators satisfy the power counting \eqref{Op}.
Whether this choice is stable under quantum corrections is the question that we are going to address in the remainder of this section. Indeed, we shall see that only a subset of operators, corresponding to very specific choices of the functions $\alpha_i$, are of WBG type and are therefore entitled to be in the group $\mathcal{O}^{\text{I}}$. The others, corresponding instead to different choices of $\alpha_i$, are required to be of type $\mathcal{O}^{\text{II}}$ and to satisfy the power counting \eqref{Op2}. 
%


For simplicity, let us focus on quantum mechanically generated interactions of type $(\nabla\phi)^{2n }$ -- the result we find will be automatically  true also for all the other couplings $c^\text{I}$ in $\mathcal{O}^\text{I}$.
In order to identify the most general WBG class, we expand the metric around a Minkowski background, $g_{\mu\nu}=\eta_{\mu\nu}+\frac{h_{\mu\nu}}{\Mpl}$, and look for theories that yield loop diagrams of type \eqref{qcH}.

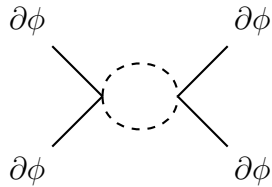
\begin{figure}[h!]
 \centering
\begin{tikzpicture}
\begin{feynman}[inline]
  \vertex (i1) {\(\partial \phi\)};
  \node[below=of i1] (n1);
  \vertex[below=of n1] (i2) {\(\partial \phi\)};  
  \vertex[right=1cm of n1] (v1);  
  \vertex[right=of v1] (v2);
  \node[right=0.85cm of v2] (n2);  
  \vertex[above=of n2] (f1){\(\partial \phi\)};  
  \vertex[below=of n2] (f2){\(\partial \phi\)};

  \diagram* {
    (i1) -- [thick,solid] (v1) -- [thick,solid] (i2),
    (v1) -- [thick,dashed,half left] (v2)
    (v1) -- [thick,dashed,half right] (v2)
    (f1) -- [thick,solid] (v2) -- [thick,solid] (f2),
  };
\end{feynman} 
\end{tikzpicture}
\caption{One-loop diagram contributing to the 2-to-2 scattering amplitude with only scalar propagators. The legs with more than one derivative are taken to be on the internal lines, represented by a dashed line.}
\label{fig:diagram-flat}
\end{figure}

To start with, let us consider the leading order in the $1/\Mpl$ expansion, i.e. we focus on loops with no internal graviton lines. The scalar self-interactions that contribute to this kind of quantum corrections are
\begin{equation}
S_{4,\phi} = \int \D^4x \left[ \frac{\Lambda_2^4}{\Lambda_3^6} C^{\mu\nu,\rho\sigma} \partial_\mu\partial_\nu\phi\partial_\rho\partial_\sigma\phi  + \ldots
\right] \, ,
\label{intS0}
\end{equation}
where we have replaced covariant derivatives with simple ones.
For a generic $C^{\mu\nu,\rho\sigma}$, the leading corrections to $(\partial\phi)^{2n }$ come from loops where there are only internal legs differentiated twice. In other words, we shall focus on the following configuration:
\begin{equation}
S_{4,\phi} = \int \D^4x \left[ \frac{\Lambda_2^4}{\Lambda_3^6} C^{\mu\nu,\rho\sigma} \partial_\mu\partial_\nu\phi_{\text{int}}\partial_\rho\partial_\sigma\phi_{\text{int}}  + \ldots
\right] \, .
\label{intloMpl}
\end{equation}
Without any prescription on $C^{\mu\nu,\rho\sigma}$, at quantum level the interactions \eqref{intloMpl} generates, in principle,
\begin{equation}
\sim \frac{(\partial\phi)^{2n}}{\Mpl^{n-2}\Lambda_3^{3n-4}} \, ,
\label{noloop}
\end{equation}
which are parametrically larger than \eqref{qcH}. On the other hand, the corrections \eqref{noloop} are not generated if $C^{\mu\nu,\rho\sigma}$ is antisymmetric under single exchange of indices between the first and the second pair:
\begin{equation}
C^{\mu\nu,\rho\sigma} = - C^{\rho\nu,\mu\sigma} \, .
\label{antsymmc}
\end{equation}
One can easily show that this occurs in theories such that
\begin{equation}
\alpha_1=-\alpha_2 \, ,
\qquad
\alpha_3=-\alpha_4 \, ,
\qquad
\alpha_5 = 0 \, ,
\qquad
\text{for all } X \, .
\label{galconds2}
\end{equation}
Notice that the conditions \eqref{galconds2} are not only sufficient but also necessary in order to guarantee that the estimation  \eqref{qcH} is correct at any order in the number of loops and for an arbitrary choice of the external momenta. To clarify this point, we shall compute for instance the amplitude associated with a $2$-to-$2$ scattering at one loop using the interaction vertices \eqref{intloMpl} where we take $\alpha_i=\text{constant}$ for $i=1,2,3,4$ and $\alpha_5=0$, as shown in Fig.~\ref{fig:diagram-flat}.
After summing over all possible permutations of the external momenta and using the symmetry properties of the loop integral, the total amplitude takes on the following form:
\begin{multline}
\mathcal{A}_{2-2}^{1-\text{loop}} \propto \left[(k_1\cdot k_2)(k_3\cdot k_4)+ (k_1\cdot k_3)(k_2\cdot k_4) + (k_1\cdot k_4)(k_2\cdot k_3) \right]
\\
\times \left[8(\alpha_1+\alpha_2)^2 + 4(\alpha_1+\alpha_2)(\alpha_3+\alpha_4)+ (\alpha_3+\alpha_4)^2 \right]
 \int\frac{\D^4q}{(2\pi)^4} \, q^4 + \ldots 
 \label{loopex1}
\end{multline}
where in the dots we are dropping terms that are higher order in the external momenta and therefore do not renormalize $(\partial\phi)^4$. The coefficient in the second line in \eqref{loopex1} is specific of the particular process under consideration: computing the same amplitude at different loop order yields a different combination of the factors $(\alpha_1+\alpha_2)$ and $(\alpha_3+\alpha_4)$. Then it is clear that in order for the amplitude to be zero at any loop order we are left with the conditions \eqref{galconds2}. Notice that, with this specific example where $\alpha_i=\text{constant}$ for $i=1,2,3,4$ and $\alpha_5=0$ we have simply recovered  the well-known non-renormalization theorem \cite{Luty:2003vm,Trodden:2011xh} (see also \cite{Goon:2016ihr}) of the flat spacetime galileons \cite{Nicolis:2008in} (we refer to App. \ref{app:dc} for the discussion about the flat-space limit). On the other hand, the points we want to emphasize here are that \textit{i)} the non-renormalization properties \eqref{qcH} are at play for general non-constant $\alpha_i(X)$ provided the conditions \eqref{galconds2} are fulfilled and that \textit{ii)} there exists a subclass of theories with specific couplings to gravity such that the result \eqref{qcH} remains true. 
In what follows we present some intermediate steps and the final result regarding this last point, refering to App. \ref{theorem} for further details on the calculations.

Going to the next-to-leading order in the $1/\Mpl$ expansion, we now consider diagrams that involve one graviton line. It is easy to show that quantum corrections to the operators in \eqref{STT} that involve external gravitons are suppressed at least by $\Lambda_3/\Mpl$ compared to the tree level couplings, as required by the definition of the class $\mathcal{O}^{\text{I}}$ (see App. \ref{theorem} for further details). Therefore, we can focus on loop diagrams where the graviton line is internal\footnote{We do not discuss diagrams with more internal graviton lines since they are more severely suppressed by additional factors of $1/\Mpl$.}. Expanding the action \eqref{STT} linearly in $h_{\mu \nu}$, the relevant vertex reads
\begin{equation}
S_4 = \int \D^4 x \left[ \frac{h_{\rho\sigma} \partial^\tau\phi}{\Lambda_3^3}  
Z^{\mu\nu,\rho\sigma}
\partial_\mu\partial_\nu \partial_\tau\phi +\ldots \right]   \, ,
\label{STT-1g}
\end{equation}
where we defined
\begin{equation}
Z^{\mu\nu,\rho\sigma} \equiv f_X  \left(\eta^{\mu\rho}\eta^{\nu\sigma} + \eta^{\mu\sigma}\eta^{\nu\rho} -2\eta^{\mu\nu}\eta^{\rho\sigma} \right)
-C^{\mu\nu,\rho\sigma}.
\end{equation}
Notice that although in some special cases the vertex \eqref{STT-1g} cancels exactly at tree level, i.e. $Z^{\mu\nu,\rho\sigma}\equiv 0$, 
this does not occur in general for the theories admitted by \eqref{galconds2}.
Nevertheless, as we shall see now, it may occur at loop level: this will define the most general class of WBG theories and extend the findings of \cite{Pirtskhalava:2015nla}.

Focussing again on the quantum mechanically generated $(\nabla\phi)^{2n}$, after the manipulations detailed in App. \ref{theorem}, loop corrections involving one internal graviton line take schematically the form
\begin{multline}
\sim -\frac{1}{4}\int\frac{\D^4q}{(2\pi)^4} \, \partial_{\tau_1}\phi\partial_{\tau_2}\phi \,  \frac{q^{\tau_1}q^{\tau_2}}{q^2}
\left(4f_X + 2X \alpha_2+ X\alpha_3\right)
\\
\times\left[
4\alpha_3\left(q^\mu\partial_\mu\phi\right)^2 + \left(12f_X + 6X \alpha_2 -X\alpha_3 \right)q^2
\right] + \ldots \, ,
\label{loopHbHsimpl10}
\end{multline}
where $q$ is the internal momentum running over the loop and where in the dots we are dropping terms that are higher orders in the external momenta and therefore do not renormalize $(\nabla\phi)^{2n}$. From this expression we can
prove that, among the theories of type \eqref{galconds2}, the WBG subclass $\mathcal{O}^{\text{I}}$ at quadratic order in $\nabla\nabla\phi$  is identified by the condition 
\begin{equation}
4f_X + 2X \alpha_2+ X\alpha_3 =0 \, .
\label{nonrencond}
\end{equation}
In the following section we shall see how both conditions \eqref{galconds2} and \eqref{nonrencond} lead to well known classes of theories.

\section{Relation with Horndeski and beyond Horndeski}
\label{sec:HandBH}

So far we have identified the most general class of theories that are structurally robust in a well defined sense. However, as outlined in the introduction, if HD operators provide the leading contribution a second ingredient is required for the consistency of these theories: the Ostrogradsky ghost-like instability should not appear. This requirement has motivated some effort in the recent literature to identify scalar tensor theories that, in spite of having higher order interactions, still propagate $3$ degrees of freedom (one scalar and the two graviton helicities). The most general set of operators with this feature is given by the so-called degenerate higher-order scalar-tensor (DHOST) theories \cite{Langlois:2015cwa,Langlois:2015skt}. Restricted up to the quadratic order (in $\nabla \nabla \phi$) DHOST theories can be defined by specific choices of the functions $\alpha_i$ in \eqref{STT}. Notice that the conditions defining these theories are less strict than the ones we obtained in \eqref{galconds2} and \eqref{nonrencond}.  
At this point, it is useful to recall that DHOST contain as two particular examples the (shift symmetric) quartic Horndeski \cite{Horndeski:1974wa,Deffayet:2011gz} and beyond-Horndeski \cite{Gleyzes:2014dya} Lagrangians, defined respectively by
\begin{equation}
X\alpha_1=-X\alpha_2=2f_X \, , \qquad \alpha_3=-\alpha_4=0 \, ,
\qquad \text{(quartic Horndeski)}
\label{cH}
\end{equation}
and
\begin{equation}
f=\frac{1}{2} \, , \qquad \alpha_1=-\alpha_2=2\alpha_3=-2\alpha_4 \, .
\qquad \text{(quartic beyond-Horndeski)}
\end{equation}
In particular, a generic linear combination of these two kinds of theories, which can be explicitly written in the form
\begin{multline}
S_4^{\text{H}+\text{bH}} =  \frac{\Lambda_2^4}{\Lambda_3^6} \int \D^4 x\sqrt{-g} \bigg[
\Lambda_2^4 G_4(X)R + 2  G_{4X}(X) \left(
\square\phi^2 - \nabla_\mu\nabla_\nu\phi \nabla^\mu\nabla^\nu\phi
\right)
\\
- \frac{ F_4(X)}{\Lambda_2^4} \epsilon^{\alpha\mu\rho\lambda} 
{\epsilon^{\beta\nu\sigma}}_\lambda \nabla_\alpha\phi\nabla_\beta\phi \nabla_\mu\nabla_\nu\phi\nabla_\rho\nabla_\sigma\phi 
\bigg] \, ,
\label{HbH4}
\end{multline}
where we have set $f = G_4$ to be in line with the standard notation, satisfies the relations
\begin{equation}
X\alpha_1=-X\alpha_2 = 2G_{4X} + XF_4 \, ,
\qquad
\alpha_3=-\alpha_4 = 2F_4 \, .
\label{HandBH}
\end{equation}
Then, it is straightforward to realize that \eqref{HandBH} coincides exactly with our conditions \eqref{galconds2} and \eqref{nonrencond}: in other words, we found that, up to quadratic order in the second derivatives of the scalar field, the most general Lagrangian belonging to the WBG class $\mathcal{O}^\text{I}$ has the form \eqref{HbH4}.

Thus we can conclude that: \textit{i)} the leading HD operators that we identified do not propagate extra degrees of freedom, \textit{ii)} our assumption about quantum stability allowed to rediscover also theories that have higher order equations of motion, i.e. beyond Horndeski, without the need to impose any degeneracy condition, \textit{iii)} among all DHOST theories, only the subset that we identified in \eqref{galconds2} and \eqref{nonrencond} seems to be able to consistently provide order-one effects at phenomenological level around the backgrounds discussed in Sec. \ref{sec:wbgpc}. Therefore, within the regime of validity of the effective expansion and barring fine tuning, DHOST theories that do not satisfy both \eqref{galconds2} and \eqref{nonrencond}, and are therefore in the subset $\mathcal{O}^\text{II}$, are typically expected to provide subdominant contributions. We will discuss more about these points in Sec. \ref{sec:pheno}.

It is worth noticing that the Horndeski type operators, defined by Eq. \eqref{cH}, are such that $Z^{\mu\nu,\rho\sigma}\equiv 0$ identically, yielding an exact cancellation at tree level of the vertex \eqref{STT-1g}. This was the guiding principle that we used to identify a specific subset of WBG operators in \cite{Pirtskhalava:2015nla}. 
With the analysis of Sec. \ref{completeWBG} we have now been able to extend the findings of \cite{Pirtskhalava:2015nla} including theories (with $Z^{\mu\nu,\rho\sigma}\neq 0$) such that the cancellation occurs at loop level -- see Eq. \eqref{loopHbHsimpl10}.

We conclude this section recalling that, after a field redefinition of the type
\begin{equation}
g_{\mu\nu} \rightarrow \tilde{g}_{\mu\nu} = A(X,\phi) g_{\mu\nu} + B(X,\phi) \frac{\nabla_\mu\phi \nabla_\nu\phi}{\Lambda_2^4} \, ,
\label{cdt}
\end{equation}
which is a combination of a conformal and a disformal transformation \cite{Bekenstein:1992pj}, one can in principle relate Horndeski and beyond-Horndeski theories to each other \cite{Zumalacarregui:2013pma,Gleyzes:2014qga,Crisostomi:2016czh,Achour:2016rkg,Crisostomi:2016tcp}. In particular, a Lagrangian of the form \eqref{HbH4} can be in general re-defined in a theory of the same type with $\tilde F_4\equiv 0$. Nevertheless, whether the two theories are physically equivalent or not depends on the specific setup under consideration. For instance, this is not the case for theories that include specific couplings to matter or possess solutions to the background equations of motion that make the transformation \eqref{cdt} singular in certain points and therefore non-invertible. These are exactly the cases that we are going to present in the next section.

\section{Phenomenology}
\label{sec:pheno}

So far we have identified the most general WBG class of operators up to quadratic order in the second derivatives of the scalar field that, thanks to some remnant of galileon's non-renormalization theorem (that we proved in Sec. \ref{completeWBG}), satisfy the power counting $\mathcal{O}^\text{I}$.
In particular, they are defined by the conditions \eqref{galconds2} and \eqref{nonrencond} and can be always written as linear combinations of the so-called Horndeski and beyond-Horndeski operators, turning the spotlight on this particular subset of the more general class of DHOST theories \cite{Langlois:2015cwa,Langlois:2015skt} as the only phenomenologically relevant ones among all possible scalar-tensor theories that propagate three degrees of freedom. 

In the following we discuss two different examples that make an EFT involving operators with WBG symmetry particularly interesting form a phenomenological point of view. 

\subsection{Naturalness in the Effective Theory of Dark Energy}

The recent combination of GW170817 \cite{TheLIGOScientific:2017qsa} and GRB170817A \cite{Goldstein:2017mmi} provides a new cornerstone in physics. Indeed, the simultaneous observation of gravitational waves and electromagnetic radiation from a single astrophysical source allowed to set strong bounds to the graviton's speed of propagation $\ct$, which has been measured to be compatible with the speed of light with deviations at most of the order of $10^{-15}$ \cite{Monitor:2017mdv}. In the context of Dark Energy models, where e.g. a scalar condensate is responsible for the current accelerated expansion of the Universe spontaneously breaking Lorentz symmetries, this is not a condition that can be given a priori. On the contrary, the observational bound on the speed of gravitational waves is reflected into severe constraints \cite{Lombriser:2015sxa,Creminelli:2017sry,Baker:2017hug,Ezquiaga:2017ekz,Sakstein:2017xjx} on the couplings in the EFT of DE \cite{Creminelli:2008wc,Gubitosi:2012hu}. For instance, in the context of shift symmetric theories, the only operators that are compatible with $\ct=1$ have been shown to be precisely those given in Eqs. \eqref{S2}, \eqref{S3} and \eqref{HbH4}, where $G_4$ and $F_4$ must satisfy in addition  $2G_{4X}=XF_4$ \cite{Creminelli:2017sry,Baker:2017hug,Ezquiaga:2017ekz,Sakstein:2017xjx}. Without any symmetry or non-renormalization property at play in general one expects order one quantum corrections to the couplings to spoil this condition and all the operators that have formally been set to zero in the tree level Lagrangian to be generated at loop level, inducing sizeable deviations from $\ct=1$.
In other words, either higher derivative operators are always phenomenologically subdominant on cosmological scales or one could not trust $\ct=1$ without a fine tuning assumption. In the present work, we have actually shown that a WBG theory is able to reconcile these two aspects: not only the operators \eqref{STT} can be as relevant as the ones in \eqref{S2} on the background, but also, thanks to the properties proved in Sec. \ref{completeWBG}, the choice $\ct=1$ is protected against large quantum corrections. Therefore the relations found by \cite{Creminelli:2017sry,Baker:2017hug,Ezquiaga:2017ekz,Sakstein:2017xjx} in light of the events GW170817 and GRB170817A do not represent a fine tuning in the theory and could be a theoretically consistent explanation to the current observations.
Indeed, choosing $\Lambda_2\sim(\Mpl H_0)^{1/2}$ and $\Lambda_3\sim(\Mpl H_0^2)^{1/3}$ to have sizeable dark energy effects, according to Tab. \ref{pctable}, we can estimate
\begin{equation}
\frac{\delta c^\text{I}}{c^\text{I}} \sim \left(\frac{H_0}{\Mpl}\right)^{2/3} \sim 10^{-40} \, ,
\end{equation}
which is far below the current sensitivity of the measurement of $\ct$. Operators in the EFT that induce deviations from $\ct=1$ are generated at the quantum level but suppressed by scales that are a few orders of magnitude larger than the ones associated with the operators that respect $\ct=1$, which are therefore the only physically relevant ones.

\subsection{Existence and stability of geodesically complete cosmologies}

Inflation is known to be past-incomplete \cite{Borde:2001nh}. Roughly speaking this means that going backward in time one has to face a singularity, i.e.  a UV completion to General Relativity is unavoidably required in order to explain the high energy regime at early times. This has motivated the search for alternative cosmologies that, relying on a violation of the Null Energy Condition (NEC), are geodesically complete. Whilst being possible to find solutions to the background equations of motion that describe genesis evolutions \cite{Creminelli:2010ba} and bounces (for a review see e.g. \cite{Novello:2008ra,Brandenberger:2016vhg}), the Lagrangian for perturbations in these models typically display, at some moment in the cosmological history, a gradient instability and/or strong coupling -- see e.g. \cite{Nicolis:2009qm,Libanov:2016kfc, Kobayashi:2016xpl,Pirtskhalava:2014esa, Easson:2011zy,Ijjas:2016tpn,Ijjas:2016vtq,Dobre:2017pnt}. Albeit many no-go examples, a systematic study of the origin of the instability and a comprehensive classification of the possible healthy theories were still lacking so far. In this context, the EFT for single-field FLRW cosmologies \cite{Creminelli:2006xe,Cheung:2007st} has been proven to be particularly useful \cite{Creminelli:2016zwa,Cai:2016thi} to fill this gap. In particular, it has been shown independently in \cite{Creminelli:2016zwa,Cai:2016thi} that the inclusion of beyond-Horndeski operators in the Lagrangian is sufficient to make the geodesically complete trajectory stable at the level of perturbations. In this case, what makes the theory non-redefinable to Horndeski through the transformation \eqref{cdt} is the fact that the solution intersects a singular point, making \eqref{cdt} non-invertible \cite{Creminelli:2016zwa}. This represents an example of physically inequivalent theories that can not be mapped to each other by conformal/disformal transformations (see \cite{Cai:2017dyi,Kolevatov:2017voe} for explicit covariant formulations).

In this context, the present work goes in the direction of supporting the reliability of this kind of theories, which seem to play a crucial role in order to have stable NEC violating cosmologies.
Without the non-trivial quantum properties that these theories manifest according to our previous discussion, it would be hard to trust a solution resulting from the tuning of the higher derivative operators in the EFT to give a geodesically complete trajectory. Indeed, as we have already emphasized, in the absence of the non-renormalization theorem, either the higher derivative operators give only subleading corrections (which cannot be the case if the NEC-violating solution has to be stabilized) or infinitely many terms in the derivative expansion are expected to equally contribute invalidating the calculability of any result in the theory.

\section{Conclusions and outlook}
\label{sec:concl}

In this work we have explicitly constructed an EFT for a scalar field coupled to gravity that contains a set of HD operators which give the dominant contribution to physical observables, at least around some nontrivial backgrounds. 
For this to be consistent, the theory must exhibit some non-renormalization property that prevents large quantum corrections to spoil the power counting of the different operators: this peculiar behavior is provided by the presence of an approximate global symmetry, which is therefore a crucial ingredient for a robust HD theory. A second condition, if HD interactions can be large, is the absence of the Ostrogradsky ghost-ljke instability below the UV cutoff of the EFT. We have derived the most general (up to quadratic order in second derivatives) interactions with these properties, which turn out to be a particular subset of quadratic DHOST theories. According to our analysis, this subset, which contains a linear combination of the so-called Horndeski (H) and beyond-Horndeski (bH) Lagrangians, is the only one where HD play a significant role when quantum corrections are included.

In the last section we have briefly discussed some implications for models of Dark Energy and for the physics of the early Universe. Mostly motivated by these phenomenological applications, in constructing the WBG class of theories we only focussed on operators that are at most quadratic in the second derivatives of the scalar field. However, following the logic that we have proposed, in principle one could straightforwardly extend the result up to cubic order and identify the corresponding WBG theory. 
It would be interesting to show whether also at this order the requirement of quantum stability is powerful enough to enforce the degeneracy condition of cubic DHOST theories \cite{BenAchour:2016fzp}, without the need of extra assumptions. This is not obvious a priori because it is known that when both quadratic and cubic H + bH are present together the degeneracy is broken and therefore the ghost propagates again \cite{Langlois:2015cwa,Crisostomi:2016tcp}. This analysis is left for future work.

Throughout the construction presented in the paper, we never made any specific assumptions about the couplings $c^\text{I}$. On the other hand, it is well known \cite{Adams:2006sv} that, if the underlying UV completion is Lorentz invariant, local and causal, valuable information about the sign of $c^\text{I}$ can be inferred from the analytic properties of the $S$-matrix. Moreover, in certain cases (e.g. galileon and massive gravity) the analysis can be extended beyond positivity constraints and employed to derive bounds on the cutoff scale or on the effective couplings \cite{Nicolis:2009qm,deRham:2017imi,Bellazzini:2017fep}. 
Since known results cannot be straightforwardly applied to the case of scalar theories coupled to a massless graviton (see \cite{Bellazzini:2017fep}), it would be interesting to look for a generalization of these techniques and see whether in WBG theories one can infer additional constraints, for instance on the parametric separation between the scales $\Lambda_2$ and $\Lambda_3$.

\subsection*{Acknowledgements}
It is a pleasure to thank B. Bellazzini, P. Creminelli, D. Pirtskhalava and F. Vernizzi for useful discussions and J. Noller for comments on the draft. We are especially thankful to Enis Belgacem for collaboration in the early stages of this project.
L.S. is supported by the Netherlands organization for scientific research (NWO). E.T. is partially supported by MIUR PRIN project 2015P5SBHT.

\appendix

\section{Decoupling limit} \label{app:dc}

The results that we have presented in the main text and derived in App. \ref{theorem} are general and do not rely on the specific form of the functions $\alpha_i(X)$ in \eqref{STT-2}. In particular, they hold irrespective of the existence of any well defined decoupling limit, i.e. a regime where the mixing with the metric can be safely neglected.
Nevertheless, it may be useful to understand what the scalar theory looks like if there exists a limit in which gravity can be turned off. 

In this section, we will assume that the functions $\alpha_i(X)$ are analytic around $X=0$ and we will consider the limit
\begin{equation}
\Mpl \rightarrow \infty \, , 
\qquad
\Lambda_2 \rightarrow \infty \, , 
\qquad
\Lambda_3 = \text{constant} \, ,
\label{dclimit}
\end{equation}
in the WBG class. In particular we expect to recover the flat-space galileon Lagrangian of \cite{Nicolis:2008in} where the galileon symmetry $\phi\rightarrow\phi+c+b_\mu x^\mu$ is exactly recovered and the non-renormalization theorem becomes an exact statement, namely quantum corrections to the couplings of $(\partial\phi)^{2n}$ are zero identically while only the trivially galileon invariant operators $\partial^m(\partial^2\phi)^n$ get renormalized.

Let us start from the general action \eqref{STT}. In the limit \eqref{dclimit} it becomes
\begin{multline}
S^{\text{dec. limit}}_{4,\phi} = \int \D^4 x \frac{1}{\Lambda_3^6} \big[
\alpha_{1}(\partial\phi)^2 \partial_\mu\partial_\nu\phi\partial^\mu\partial^\nu\phi + \alpha_{2}(\partial\phi)^2(\square\phi)^2 
\\
+ \alpha_3\partial^\mu\phi\partial^\nu\phi\partial_\mu\partial_\nu\phi  \square\phi
+ \alpha_4 \partial^\mu\phi\partial^\nu\phi\partial_\mu\partial_\rho\phi\partial_\nu\partial^\rho\phi
\big] \, ,
\label{STTdc}
\end{multline}
where now the functions $\alpha_i$ are computed at $X=0$.
After simple integrations by parts,
\begin{multline}
S^{\text{dec. limit}}_{4,\phi} = \int \D^4 x \frac{1}{\Lambda_3^6} \bigg[
\left(\alpha_{1}-\frac{\alpha_4}{2} \right)(\partial\phi)^2 \partial_\mu\partial_\nu\phi\partial^\mu\partial^\nu\phi
\\
+ \left(\alpha_{2}-\frac{\alpha_3}{2}\right)(\partial\phi)^2(\square\phi)^2 
-\frac{1}{2}(\alpha_3+\alpha_4)(\partial\phi)^2\partial^\mu\phi\partial_\mu\square\phi
\bigg] \, .
\label{STTdc-2}
\end{multline}
Then it is clear that the quartic galileon Lagrangian \cite{Nicolis:2008in} is recovered if
\begin{equation}
\alpha_1(0)=-\alpha_2(0) \, ,
\qquad
\alpha_3(0)=-\alpha_4(0) \, ,
\label{galconds}
\end{equation}
at the leading order in the expansion around $X=0$. Notice that Eqs. \eqref{galconds} are a particular case of \eqref{galconds2}, fixing the value of the $\alpha_i$'s in the single point $X=0$.

We conclude stressing again that, not only the conditions \eqref{galconds2} turn out to be stronger than those in Eq. \eqref{galconds}, but they are also more general in the sense that they do not rely on any assumption about the expandability around $X=0$. The results of Sec. \ref{completeWBG} remain true even if there is no limit in which the standard flat-space galileons are recovered.

\section{The non-renormalization theorem} \label{theorem}

In this appendix we derive explicitly the condition \eqref{nonrencond}, which, together with \eqref{galconds2}, defines the operators of type \eqref{STT} that belong to the class $\mathcal{O}^\text{I}$.  In particular we are interested in the next-to-leading order in the expansion $g_{\mu\nu}=\eta_{\mu\nu}+\frac{h_{\mu\nu}}{\Mpl}$, since we have already discussed the exactly flat spacetime limit in the main text.

First, we show that, if \eqref{antsymmc} holds, quantum corrections to the operators in \eqref{STT} involving external gravitons are always suppressed by $\Lambda_3/\Mpl$, in agreement with the corresponding power counting of Tab. \ref{pctable}\footnote{We thank P. Creminelli, M. Lewandowski, G. Tambalo and F. Vernizzi for a nice discussion related to this point.}. To this end, it is useful to re-write the second term in \eqref{STT} as
\begin{equation}
\begin{split}
C^{\mu\nu,\rho\sigma}\nabla_\mu \nabla_\nu \phi\nabla_\rho \nabla_\sigma \phi 
& = - (\nabla_\rho C^{\mu\nu,\rho\sigma})\nabla_\mu \nabla_\nu \phi \nabla_\sigma \phi
- C^{\mu\nu,\rho\sigma}{R^\lambda}_{\nu\rho\mu} \nabla_\lambda \phi \nabla_\sigma \phi 
\\
&	=  (\nabla_{[\mu} \nabla_{\rho]} C^{\mu\nu,\rho\sigma}) \nabla_\nu \phi \nabla_\sigma \phi
- C^{\mu\nu,\rho\sigma}{R^\lambda}_{\nu\rho\mu} \nabla_\lambda \phi \nabla_\sigma \phi \, ,
\end{split}
\end{equation}
where we integrated the covariant derivatives by parts and used the antisymmetry condition \eqref{antsymmc}. This makes transparent that all the operators in \eqref{STT} satisfying \eqref{antsymmc} can be recast in a form in which they are linear in the Riemann tensor. Since it contains two derivatives acting on the graviton line, it is easy now to understand why loop corrections are suppressed by at least $\Lambda_3/\Mpl$.

Then, we shall focus on quantum mechanically generated loop diagrams that involve one graviton internal line. These will provide the leading corrections to the couplings $\delta c^\text{I} $: diagrams with more internal gravitons are more severely suppressed by higher powers of $1/\Mpl$.
We start with some useful formulae: expanding up to linear order in $h_{\mu\nu}$, we find
\begin{equation}
\nabla_\rho\nabla_\sigma\phi = -\frac{1}{2\Mpl}\left(\partial_\rho h_{\tau\sigma} + \partial_\sigma h_{\tau\rho} -
\partial_\tau h_{\rho\sigma}\right) \partial^\tau\phi + \ldots 
\end{equation}
\begin{equation}
\sqrt{-g }f R = \frac{2 f_X}{\Lambda_2^4} \frac{h^{\rho\sigma}}{\Mpl} \partial^\tau\phi \left( \partial_\rho\partial_\sigma-\eta_{\rho\sigma}\square \right)\partial_\tau\phi + \ldots 
\end{equation}
where in the dots we are dropping total derivatives, terms with fewer factors of $\partial\phi$ and higher orders in $1/\Mpl$. Then,
\begin{equation}
C^{\mu\nu,\rho\sigma} \nabla_\mu\nabla_\nu\phi\nabla_\rho\nabla_\sigma\phi 
= \frac{1}{\Mpl}C^{\mu\nu,\rho\sigma}\partial^\tau\phi
\left( h_{\tau\sigma}\partial_\rho +  h_{\tau\rho} \partial_\sigma -
 h_{\rho\sigma}\partial_\tau\right) \partial_\mu\partial_\nu\phi + \ldots
\label{Cexp1}
\end{equation}
Using the antisymmetry conditions \eqref{antsymmc}, Eq. \eqref{Cexp1} simply reads
\begin{equation}
C^{\mu\nu,\rho\sigma} \nabla_\mu\nabla_\nu\phi\nabla_\rho\nabla_\sigma\phi 
= -\frac{1}{\Mpl}C^{\mu\nu,\rho\sigma}h_{\rho\sigma}\partial^\tau\phi
 \partial_\tau \partial_\mu\partial_\nu\phi + \ldots
\label{Cexp2}
\end{equation}
Plugging into \eqref{STT}, 
\begin{equation}
S_4 = \int \D^4 x \left[ \frac{h_{\rho\sigma} \partial^\tau\phi}{\Lambda_3^3}  
Z^{\mu\nu,\rho\sigma}
\partial_\mu\partial_\nu \partial_\tau\phi +\ldots \right]   \, ,
\label{STT-1g-2}
\end{equation}
where we defined
\begin{equation}
Z^{\mu\nu,\rho\sigma} \equiv f_X  \left(\eta^{\mu\rho}\eta^{\nu\sigma} + \eta^{\mu\sigma}\eta^{\nu\rho} -2\eta^{\mu\nu}\eta^{\rho\sigma} \right)
-C^{\mu\nu,\rho\sigma}.
\end{equation}
As mentioned in the main text, for the Horndeski class ($\alpha_1=-\alpha_2=2f_X/X$, $\alpha_3=-\alpha_4=0$) the vertex \eqref{STT-1g-2} cancels exactly at tree level, i.e. $Z^{\mu\nu,\rho\sigma}\equiv 0$ \cite{Pirtskhalava:2015nla}. For other theories admitted by \eqref{galconds2}, including beyond-Horndeski ($f=1/2$, $\alpha_1=-\alpha_2=2\alpha_3=-2\alpha_4$), there may be instead a cancellation at loop level, defining the most general class of WBG theories.

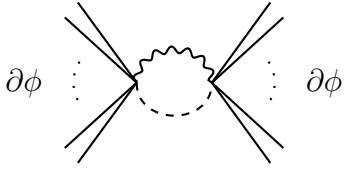
\begin{figure}[h!]
 \centering
\begin{tikzpicture}
\begin{feynman}[inline]
  \vertex (i1) {};
  \vertex[below left=0.3cm of i1] (i2) {};
  \vertex[below=of i2] (i3) {};
  \vertex[below=of i3] (i4) {};
  \vertex[below right=0.3cm of i4] (i5) {};
  \vertex[right=1.1cm of i3] (v1);  
  \vertex[right=of v1] (v2);
  \vertex[right=0.95cm of v2] (f3) {};  
  \vertex[above=of f3] (f2) {};
  \vertex[above left=0.3cm of f2] (f1) {};
  \vertex[below=of f3] (f4) {};  
  \vertex[below left=0.3cm of f4] (f5) {};

  \node[above right=0.6cm of i3] (n1);
  \node[below right=0.6cm of i3] (n2);
  \node[above left=0.6cm of f3] (m1);
  \node[below left=0.6cm of f3] (m2);
  \node[left=0.4cm of i3] (l1) {\(\partial \phi\)};
  \node[right=0.4cm of f3] (l2) {\(\partial \phi\)};
  
  \diagram* {
    (i1) -- [thick,solid] (v1),
    (i2) -- [thick,solid] (v1),
    (i4) -- [thick,solid] (v1),
    (i5) -- [thick,solid] (v1),
    (v1) -- [thick,photon,half left] (v2)
    (v1) -- [thick,dashed,half right] (v2)
    (f1) -- [thick,solid] (v2),
    (f2) -- [thick,solid] (v2),
    (f4) -- [thick,solid] (v2),
    (f5) -- [thick,solid] (v2),
    (n1) -- [thick,loosely dotted,bend right] (n2),
    (m1) -- [thick,loosely dotted,bend left] (m2),
  };
\end{feynman} 
\end{tikzpicture}
\caption{One-loop diagram with a single internal graviton line that potentially gives large quantum corrections to the operator $(\partial \phi)^{2n}$. Again the internal dashed line represents the scalar legs with more than one derivative.}
\label{fig:loop-graviton}
\end{figure}

As an illustrative example, let us focus on the loop generated operators $(\nabla\phi)^{2n}$.
The leading corrections are provided by loop diagrams involving two vertices of type  \eqref{STT-1g}\footnote{It is easy to show that loop diagrams involving one vertex of type \eqref{STT-1g} and one vertex of type \eqref{S3} are identically zero if \eqref{galconds2} holds, leading automatically to the estimation in Tab. \ref{pctable}.} with $h_{\rho\sigma}$ and $\partial^3\phi$ as internal lines (see Fig.~\ref{fig:loop-graviton}).
%
Unless some cancellations of the types advocated above occur, by simple dimensional analysis, such quantum corrections are in the form \eqref{noloop} and do not generically fit in the WBG class.
However, it is possible to find a general condition such that these unwanted corrections turn out to be identically zero, making the estimation \eqref{Op} the dominant contribution and defining therefore the most general WBG theory at quadratic order in $\nabla\nabla\phi$.
Using
\begin{equation}
\mathcal{D}_{\mu\nu,\rho\sigma}(q^2) = \frac{-i}{2q^2} \left(\eta_{\mu\rho}\eta_{\nu\sigma} + \eta_{\mu\sigma}\eta_{\nu\rho} - \eta_{\mu\nu}\eta_{\rho\sigma} \right) 
\end{equation}
for the massless graviton propagator in $(3+1)$-dimensions, the loop integral in Fig.~\ref{fig:loop-graviton} takes on schematically the form
\begin{multline}
\sim  \int\frac{\D^4q}{(2\pi)^4} \, \partial_{\tau_1}\phi\partial_{\tau_2}\phi \,  \frac{q_{\alpha_1}q_{\alpha_2}q_{\beta_1}q_{\beta_2}q^{\tau_1}q^{\tau_2}}{q^4}
Z^{\alpha_1\beta_1,\rho_1\sigma_1}Z^{\alpha_2\beta_2,\rho_2\sigma_2}
\\
\times\left(\eta_{\rho_1\rho_2}\eta_{\sigma_1\sigma_2} + \eta_{\rho_1\sigma_2}\eta_{\sigma_1\rho_2} - \eta_{\rho_1\sigma_1}\eta_{\rho_2\sigma_2} \right)
+ \ldots
 \, ,
\label{loopHbH}
\end{multline}
where $q$ is the internal momentum running over $\partial^3\phi$ and where in the dots we are dropping terms that are higher orders in the external momenta and therefore do not renormalize $(\nabla\phi)^{2n}$.
Notice that depending on the number of internal legs and the loop order there might be multiple integrals in \eqref{loopHbH}\footnote{Additional external/internal momenta have been left unexpressed in $Z^{\alpha\beta,\rho\sigma}(X)$.}. In other words, the loop correction will in general look more complicated than the schematic expression that we have reported in \eqref{loopHbH}. Nevertheless, for our purposes, only the index contractions among the $Z$-tensors and the graviton propagator, that we have highlighted in \eqref{loopHbH}, turn out to be relevant.
Indeed, after straightforward algebraic simplifications, Eq. \eqref{loopHbH} becomes
\begin{multline}
\sim -\frac{1}{4}\int\frac{\D^4q}{(2\pi)^4} \, \partial_{\tau_1}\phi\partial_{\tau_2}\phi \,  \frac{q^{\tau_1}q^{\tau_2}}{q^2}
\left(4f_X + 2X \alpha_2+ X\alpha_3\right)
\\
\times\left[
4\alpha_3\left(q^\mu\partial_\mu\phi\right)^2 + \left(12f_X + 6X \alpha_2 -X\alpha_3 \right)q^2
\right] + \ldots \, .
\label{loopHbHsimpl1}
\end{multline}
Notice that the relative coefficients that enter the combination in the second line of \eqref{loopHbHsimpl1} will depend on the the specific loop diagram under consideration, while the term in parenthesis in the the first line always provides the same overall factor. 
Therefore, the loop \eqref{loopHbHsimpl1} is identically zero if and only if
\begin{equation}
4f_X + 2X \alpha_2+ X\alpha_3 =0 \, ,
\label{finres4}
\end{equation}
which is condition \eqref{nonrencond}.

\bibliographystyle{utphys}
\bibliography{BH-nonren}

\end{document}